\documentclass[aps,pra,twocolumn,superscriptaddress]{revtex4-1}
\usepackage{bm}
\usepackage{graphicx}
\usepackage{amssymb,amsmath,amsbsy,amsgen,amsfonts}
\usepackage{dcolumn}
\usepackage{amsthm}
\usepackage{mathrsfs}
\usepackage{latexsym}
\usepackage{array}
\usepackage{amstext}
\usepackage{epsfig}
\usepackage{epstopdf} 

\usepackage{color}
\usepackage{units}
\usepackage{calrsfs}
\usepackage{verbatim}
\DeclareMathAlphabet\mathbfcal{OMS}{cmsy}{b}{n}
\usepackage{mathtools}

\newcommand{\be}{\begin{equation}}
\newcommand{\ee}{\end{equation}}
\newcommand{\ba}{\begin{array}}
\newcommand{\ea}{\end{array}}
\newcommand{\bqa}{\begin{eqnarray}}
\newcommand{\eqa}{\end{eqnarray}}

\renewcommand{\vec}[1]{\bm{#1}}
\newcommand{\tens}[1]{\mathbf{#1}}

\newcommand{\sprod}{\!\cdot\!}
\newcommand{\tprod}{}

\newcommand{\dif}{\mathrm{d}}
\newcommand{\mi}{\mathrm{i}} 
\newcommand{\me}{\mathrm{e}}

\begin{document}

\title{
The Langevin Noise Approach for Lossy Media and the Lossless Limit}

\author{George W. Hanson}
\email[]{george@uwm.edu}
\address{Department of Electrical Engineering, University of Wisconsin-Milwaukee, 3200 N. Cramer St., Milwaukee, Wisconsin 53211, USA}

\author{Frieder Lindel}
\address{Physikalisches Institut, Albert-Ludwigs-Universit\"{a}t Freiburg, Hermann-Herder-Stra\ss e 3, 79104 Freiburg, Germany}

\author{Stefan Yoshi Buhmann}
\email[]{stefan.buhmann@physik.uni-freiburg.de}
\address{Physikalisches Institut, Albert-Ludwigs-Universit\"{a}t Freiburg, Hermann-Herder-Stra\ss e 3, 79104 Freiburg, Germany}

\date{\today }

\begin{abstract}
The Langevin noise approach for quantization of macroscopic
electromagnetics for three-dimensional, inhomogeneous environments is compared
with normal mode quantization. Recent works on the applicability of the method
are discussed, and several examples are provided showing that for closed
systems the Langevin noise approach reduces to the usual cavity mode expansion method when loss is eliminated.

\end{abstract}

\maketitle

\section{Introduction}

Methods for the study of the quantum properties of light, and the interaction
of quantized light and atoms and other multi-leveled systems, were initially
developed for vacuum. The observation of Purcell in 1946 that the spontaneous
emission rate of an atom was dependent on the atom's environment
\cite{Purcell} was a motivating factor for the study of how cavity materials
affect quantized light. The incorporation of simplified models of materials
(lossless, dispersionless dielectrics, perfect metals) is accommodated in
quantum models in a fairy straightforward manner \cite{DM}. 
However, the Kramers-Kronig
relations \cite{Jac} require that absorption is always accomplished by
dispersion, and vice versa. Whereas in classical electromagnetics, dispersion
and absorption are easily accounted for, in macroscopic quantum models this is
not the case, since a naive implementation of absorption causes the
commutators to vanish at long times, violating the Heisenberg uncertainty principle.

Motivated by the fluctuation--dissipation theorem,
\cite{HB}-\cite{B2}---describe macroscopic quantum electrodynamics (QED) as inspired by its nature as the quantum version of classical macroscopic electrodynamics---is 
a phenomenological dipolar, fully quantum,
macroscopic theory developed to accommodate lossy, dispersive materials and
open environments. It has been widely applied to a variety of problems since
it is expressed in terms of the Green function, and allows for very general
media, including anisotropic, nonreciprocal, and nonlocal materials
\cite{Buh}, \cite{RSW}-\cite{Force2}. For inhomogeneous, complex-shaped
regions, the Green function can be computed numerically \cite{Cole}. In
Ref.~\cite{Phil}, the phenomenological assumptions are derived from a canonical
formulation; 
this approach was later extended to moving media \cite{Horsley12}. The equivalence of the approach with an alternative based on auxiliary fields \cite{0710} was demonstrated explicitly \cite{0711}.
A critical assessment is provided in \cite{AD} (see also
\cite{AD1})-\cite{AD2}), where a comparison with a generalized Huttner-Barnett
approach \cite{HB} (canonical quantization of a bath of oscillators, based on
\cite{Hop}) is discussed. Dissipation and dielectric models are also discussed
in a wide range of other works, see e.g. \cite{Chew}.

In \cite{AD}, the practical equivalence of the Langevin noise approach (LNA) and Huttner-Barnett
descriptions is shown. More precisely, it is shown that in an open system, the
material oscillator degrees of freedom included in the standard Langevin noise approach
must be augmented by quantized photonic degrees of freedom associated with
fluctuating fields coming from infinity and scattered by the inhomogeneities
of the medium. If space is considered to consist of a uniform background
having some small absorption, the free fields coming from infinity are
absorbed and the standard Langevin noise approach applies. However, it is often of interest to
model finite regions of space having nonabsorbing materials. In \cite{AD} a
scheme is developed considering a finite region of space (which may be
vacuum), surrounded by a weakly absorbing/dispersive medium $\varepsilon
_{\text{inf}}$ that extends to infinity, and fluctuating polarization currents
in $\varepsilon_{\text{inf}}$\ generate the missing free fields, in which case
the Huttner-Barnett and Langevin noise approaches are shown to be equivalent.

Nevertheless, questions about the validity of the Langevin noise approach remain \cite{DLGJ}%
-\cite{DGJ}, particularly, concerning various limiting procedures such as
assuming the material region of interest shrinks to zero, or the limit of a
lossless material is taken. In this work, we compare the Langevin noise approach with the standard
cavity normal mode approach, which we refer to as normal-mode QED (NMQED) in the following, which is valid for media characterized by
Hermitian permittivity tensors (lossless, and therefore, nondispersive).
Although it is known that the Langevin noise approach recovers various quantities correctly, such
as the atomic spontaneous decay rate, here we show for several explicit
examples that the Langevin noise approach results in exactly the same formulation (final
equations) as the normal-mode QED, although the former allows for much more general
materials than the latter. Several possible geometries may be envisioned: 1)
finite-size, PEC-wall cavities (i.e., closed systems) containing lossless
inhomogeneous media, 2) same as (1) but for lossy, dispersive media, 3)
large-cavity limit cavities containing lossless inhomogeneous media, 4) same
as (3) but for lossy inhomogeneous media, 5) open systems, which admit loss
even when the materials themselves are lossless. Cases (3) and (4) are
actually subsets of (1) and (2); in the former, plane-wave eigenfunctions are
used, whereas in the latter, more general cavity eigenfunctions are used. For
(1), normal-mode QED is standard, often with homogeneous media (e.g., vacuum). The Langevin noise approach
does not apply to Case (1) directly, but can be applied to Case (2), the lossy
version. Here, we show that the Langevin noise approach recovers exactly the normal-mode QED equations for
several problems considered in the lossless limit (i.e., as Case (2) reduces
to Case (1)). For Case (3), the normal-mode QED is often used for homogeneous
environments (utilizing discrete plane-wave mode functions to represent the
actual mode continuum). Again, the Langevin noise approach can not be applied directly to Case
(3), although it applies to Case (4) and again recovers the normal-mode QED result in the
lossless limit. In fact, the resulting equations from the Langevin noise approach, e.g., the
density operator or population evolution, are easily converted to the normal-mode QED
(and, sometimes, vice-versa) using a simple Green function relation. For the
study of non-absorbing materials, we point out the need to retain dissipation
in the Langevin noise approach model until the final steps of the calculation, at which point the
lossless limit can be taken. Similarly, if, say, the medium inhomogeneities
vanish (e.g., the structure of interest, such as a metal resonator, shrinks to
zero size), that limit must be taken at the end of the development. 

Open systems, case (5), cannot be modeled using cavity normal modes, but it
can be modeled using Langevin noise approach (in the references cited above, it is inherently a
system-bath approach). For open systems, a quasinormal mode quantization (also
based on a Langevin noise model) is a useful and natural approach for
arbitrarily lossy open system modes \cite{SH1}, and implements a formulation
akin to the standard modal approach, but for open lossy systems. An advantage
of quasinormal modes beyond the Langevin noise approach is to explore nonlinear quantum optics at
the system level, where it is no longer valid to treat the medium as a bath,
e.g. \cite{SH2}--\cite{SH3}.

\section{Basic Relations}

We first consider an environment/reservoir such as a three-dimensional cavity
$\Omega\subseteq\mathbb{R}^{3}$ with closed surface $\Sigma$, having a uniform
background material characterized by $\varepsilon_{\text{bulk}}$ and
containing a region $\Omega_{1}\subseteq\Omega$ inhomogeneously-filled with
material characterized by relative permittivity tensor $\mathbf{\varepsilon
}_{1}\left(  \mathbf{r},\omega\right)  $ (this could be, e.g., a plasmonic
material). The permittivity for all $\mathbf{r}\in\Omega$ is
$\mathbf{\varepsilon}\left(  \mathbf{r},\omega\right)  $. We will assume the
magnetic permeability is the unity tensor, although including a permeability
response does not change the presented conclusions. As the notation indicates,
we can allow $\Omega_{1}=\Omega$, and $\Omega$ can be finite (e.g., a closed
system with surface $\Sigma$ perfectly conducting), or in the large-cavity limit. The geometry is
depicted in Fig. 1, including a two-level system located somewhere within
$\Omega$. We compare two formulations.%
\begin{figure}
  \includegraphics[width=0.7\columnwidth]{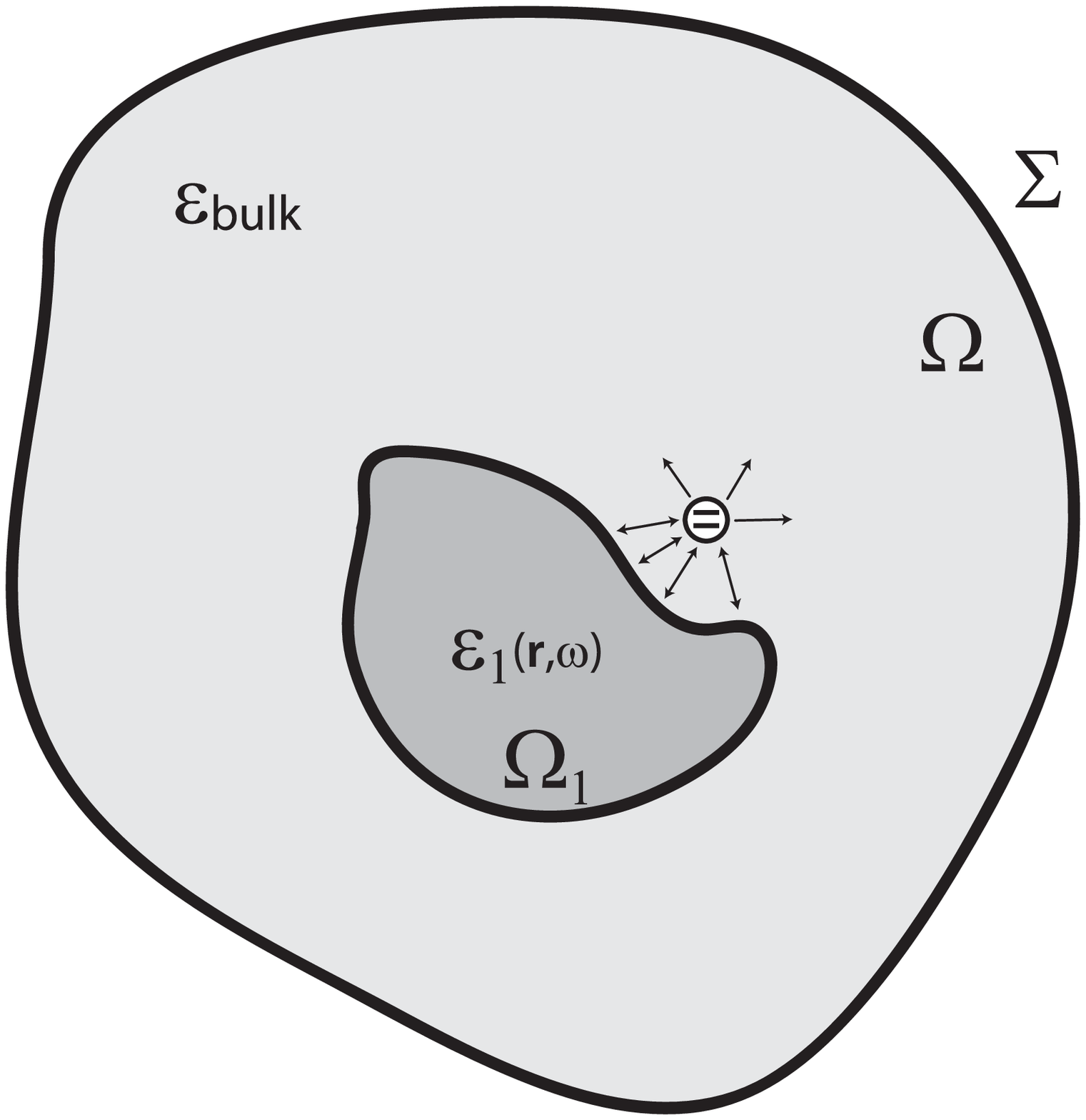}
  \caption{
Two-level system in the vicinity of an inhomogeneous region
$\Omega_{1}\subseteq\Omega\subseteq\mathbb{R}^{3}$
  }
 \label{fig1}
\end{figure}

\subsection{Normal-Mode QED Approach}

Normal-mode QED is the usual textbook \cite{QOB1}-\cite{QOB4} and research
\cite{RA1}-\cite{RA2} approach for i) closed empty cavities, where
$\mathbf{\varepsilon}\left(  \mathbf{r},\omega\right)  =\mathbf{I}$, with
$\mathbf{I}$ the identity operator, ii) closed cavities filled with lossless,
dispersionless media, where $\mathbf{\varepsilon}\left(  \mathbf{r}\right)  $
is a real-valued, Hermitian tensor, and iii) closed cavities
homogeneously-filled with lossy media. For the first two cases, classical mode
functions $\mathbf{E}_{\mathbf{k}}\left(  \mathbf{r}\right)  =\mathbf{E}%
_{\mathbf{k}}\left(  \mathbf{r},\omega_{\mathbf{k}}\right)  $ can be defined
that satisfy \cite{VanB}, \cite{Wubs2}%
\begin{equation}
\nabla\times\nabla\times\mathbf{E}_{\mathbf{k}}\left(  \mathbf{r}%
,\omega_{\mathbf{k}}\right)  =\frac{\omega_{\mathbf{k}}^{2}}{c^{2}%
}\mathbf{\varepsilon}\left(  \mathbf{r}\right)  \cdot\mathbf{E}_{\mathbf{k}%
}\left(  \mathbf{r},\omega_{\mathbf{k}}\right)  ,\label{efe}%
\end{equation}
subject to boundary conditions on the cavity walls, $\left.  \widehat
{\mathbf{n}}\left(  \mathbf{r}\right)  \times\mathbf{E}_{\mathbf{k}}\left(
\mathbf{r},\omega_{\mathbf{k}}\right)  \right\vert _{\mathbf{r}_{\text{wall}}%
}=\mathbf{0}$, $\widehat{\mathbf{n}}$ being the unit normal vector to the
wall, with eigenfunction orthogonality \cite{VanB}
\begin{equation}
\int\mathbf{E}_{\mathbf{k}}^{\ast}\left(  \mathbf{r},\omega_{\mathbf{k}%
}\right)  \cdot\mathbf{\varepsilon}\left(  \mathbf{r}\right)  \cdot
\mathbf{E}_{\mathbf{k}^{\prime}}\left(  \mathbf{r},\omega_{\mathbf{k}^{\prime
}}\right)  d^{3}\mathbf{r}=\delta_{\mathbf{kk}^{\prime}}.
\end{equation}
Under the restriction of a Hermitian permittivity tensor, and defining the
Hilbert space of Lebesgue integrable vector functions $\mathbf{L}^{2}$, the
operator $L_{E}:\mathbf{L}^{2}\left(  \Omega\right)  ^{3}\rightarrow
\mathbf{L}^{2}\left(  \Omega\right)  ^{3}$, $L_{E}\mathbf{x}\equiv
\mathbf{\nabla\,\times\,\nabla}\times\mathbf{x}-\frac{\omega_{\mathbf{k}}^{2}%
}{c^{2}}\mathbf{\varepsilon}(\mathbf{r})\cdot\mathbf{x}$, with boundary
condition $B\left(  \mathbf{x}\right)  =\left.  \mathbf{n\times x}\right\vert
_{\Gamma}\mathbf{=0}$ or $B\left(  \mathbf{x}\right)  =\left.  \mathbf{n\times
\nabla\times x}\right\vert _{\Gamma}\mathbf{=0}$ is self adjoint (SA) and
negative-definite, and the modes form an orthonormal, complete set in the
Hilbert space of square-integrable functions \cite{VanB},%
\begin{equation}
\mathbf{I}\delta\left(  \mathbf{r}-\mathbf{r}^{\prime}\right)  =%
{\displaystyle\sum\nolimits_{\mathbf{k}}}
\mathbf{E}_{\mathbf{k}}\left(  \mathbf{r},\omega_{\mathbf{k}}\right)
\mathbf{E}_{\mathbf{k}}^{\ast}\left(  \mathbf{r}^{\prime},\omega_{\mathbf{k}%
}\right)  \cdot\mathbf{\varepsilon}\left(  \mathbf{r}^{\prime}\right)  .
\end{equation}

The electric field operator in the Schr\"{o}dinger picture is%
\begin{equation}
\widehat{\mathbf{E}}\left(  \mathbf{r}\right)  ^{\text{NMQED}}=%
{\displaystyle\sum\nolimits_{\mathbf{k}}}
\widehat{\mathbf{E}}_{\mathbf{k}}\left(  \mathbf{r}\right)  +\text{H.c.}%
,\label{EFO1}%
\end{equation}
where%
\begin{equation}
\widehat{\mathbf{E}}_{\mathbf{k}}\left(  \mathbf{r}\right)  =i\sqrt
{\frac{\hslash\omega_{\mathbf{k}}}{2\varepsilon_{0}}}\widehat{a}_{\mathbf{k}%
}\mathbf{E}_{\mathbf{k}}\left(  \mathbf{r}\right) \label{EFO2}%
\end{equation}
and where $\widehat{a}_{\mathbf{k}},\widehat{a}_{\mathbf{k}}^{\dag}$ are
annihilation and creation operators that satisfy
\begin{equation}
\left[  \widehat{a}_{\mathbf{k}},\widehat{a}_{\mathbf{k}^{\prime}}\right]
=\left[  \widehat{a}_{\mathbf{k}}^{\dagger},\widehat{a}_{\mathbf{k}^{\prime}%
}^{\dagger}\right]  =0,\ \left[  \widehat{a}_{\mathbf{k}},\widehat
{a}_{\mathbf{k}^{\prime}}^{\dagger}\right]  =\delta_{\mathbf{kk}^{\prime}%
}.\label{sc}%
\end{equation}
In the Heisenberg picture, (\ref{sc}) become equal-time commutators.

The free-field Hamiltonian is (dropping the zero-point energy)
\begin{equation}
\widehat{H}^{\text{NMQED}}=%
{\displaystyle\sum\nolimits_{\mathbf{k}s}}
\hslash\omega_{\mathbf{k}}\widehat{a}_{\mathbf{k}s}^{\dagger}\widehat
{a}_{\mathbf{k}s},\label{CMH}%
\end{equation}
and eigenfunctions of the Hamiltonian are the multimode number (Fock) states
\begin{equation}
\left\vert n_{1}\right\rangle \left\vert n_{2}\right\rangle \left\vert
n_{3}...\right\rangle \equiv\left\vert n_{1},n_{2},n_{3}...\right\rangle
=\left\vert \left\{  n_{j}\right\}  \right\rangle ,
\end{equation}
which can be obtained from the ground state as%
\begin{equation}
\left\vert n_{1}\right\rangle \left\vert n_{2}\right\rangle ...n_{i}%
,...=\frac{\left(  \widehat{a}_{1}^{\dagger}\right)  ^{n_{1}}}{\sqrt{n_{1}!}%
}...\frac{\left(  \widehat{a}_{i}^{\dagger}\right)  ^{n_{i}}}{\sqrt{n_{i}!}%
}...\left\vert 0\right\rangle .
\end{equation}

For the special case of an optically large vacuum cavity, the cavity mode
functions become%
\begin{equation}
\mathbf{E}_{\mathbf{ks}}\left(  \mathbf{r}\right)  \rightarrow\frac
{\mathbf{e}_{\mathbf{k}s}}{\sqrt{V}}e^{i\mathbf{k}\cdot\mathbf{r}},\label{PWEF}%
\end{equation}
which satisfy periodic boundary conditions ($\Omega$ is assumed to be the
union of boxes of volume $V$), where $s$ indicates spin (polarization), with
$\mathbf{e}_{\mathbf{k}s}$ being an orthonormal set of polarization functions
such that $\mathbf{e}_{\mathbf{k}s}\cdot\mathbf{e}_{\mathbf{k}^{\prime
}s^{\prime}}=\delta_{\mathbf{kk}^{\prime}}\delta_{ss^{\prime}}$, and satisfy
the transversality condition $\mathbf{k}\cdot\mathbf{e}_{\mathbf{k}s}=0$. The
polarization vectors form a right-handed coordinate system, $\mathbf{e}%
_{\mathbf{k}1}\times\mathbf{e}_{\mathbf{k}2}=\mathbf{k/\left\vert
\mathbf{k}\right\vert }$. In (\ref{PWEF}), $V$ is a quantization volume such
that%
\begin{equation}
\int_{V}\mathbf{E}_{\mathbf{ks}}^{\ast}\left(  \mathbf{r}\right)
\cdot\mathbf{E}_{\mathbf{k}^{\prime}s^{\prime}}\left(  \mathbf{r}\right)
d^{3}\mathbf{r}=\delta_{\mathbf{kk}^{\prime}}\delta_{ss^{\prime}}.
\end{equation}
Note, however, that this is not an open system (truly infinite space), which
inherently allows dissipation (photons going to infinity and never coming
back). Mathematically, the difference between a large cavity and a true open
system is that for the latter, modes must satisfy the Sommerfeld radiation
condition, which renders the operator $L_{E}$ to be non-self-adjoint; the Sommerfeld radiation
condition is an out-going wave condition, and the adjoint condition is an
inward-traveling wave.

Finally, for case (iii), a cavity homogeneously-filled with lossy media,
rather than $L_{E}$, the operator $L\mathbf{x}\equiv\mathbf{\nabla
\,\times\,\nabla}\times\mathbf{x}$ can be defined such that eigenfunctions of
$L$ satisfy the boundary condition $B\left(  \mathbf{x}\right)  =\mathbf{0}$,
and the resulting operator is SA. The cavity must be
homogeneously-filled; material inhomogeneities in piecewise constant media
would necessitate boundary conditions $B$ such that $B\neq B^{\ast}$,
rendering the problem non-self-adjoint.

In the usual normal-mode QED, the photonic Green functions is not explicitly needed,
although it implicitly arises in, e.g., atom-atom coupling terms. However, to
make connection with the Langevin noise approach, it is important to connect the mode functions
$\mathbf{E}_{\mathbf{k}}\left(  \mathbf{r},\omega_{\mathbf{k}}\right)  $ with
the Green tensor, which is defined by
\begin{equation}
\nabla\times\nabla\times\mathbf{G}\left(  \mathbf{r},\mathbf{r}^{\prime
},\omega\right)  -\frac{\omega_{\mathbf{k}}^{2}}{c^{2}}\mathbf{\varepsilon
}\left(  \mathbf{r}\right)  \cdot\mathbf{G}\left(  \mathbf{r},\mathbf{r}%
^{\prime},\omega\right)  =\mathbf{I}\delta\left(  \mathbf{r}-\mathbf{r}%
^{\prime}\right)
\end{equation}
and satisfies $\mathbf{G(r},\mathbf{r}^{\prime}\mathbf{)}^{\top}%
=\mathbf{G(r}^{\prime},\mathbf{r)}$. The Green tensor can be expanded as
\begin{equation}
\mathbf{G}\left(  \mathbf{r},\mathbf{r}^{\prime},\omega\right)  =%
{\displaystyle\sum\nolimits_{\mathbf{k}}}
c^{2}\frac{\mathbf{E}_{\mathbf{k}}\left(  \mathbf{r},\omega_{\mathbf{k}%
}\right)  \mathbf{E}_{\mathbf{k}}^{\ast}\left(  \mathbf{r}^{\prime}%
,\omega_{\mathbf{k}}\right)  }{\omega_{\mathbf{k}}^{2}-\omega^{2}}.\label{GTE}%
\end{equation}
The expression (\ref{GTE}) formally encompasses the case of transverse modes,
forming a transverse Green function, or could include longitudinal modes as
well. It should be emphasized that (\ref{GTE}) is only valid for closed
cavities and the three cases discussed, although the Green tensor concept
itself extends to dispersive and lossy inhomogeneous media. For certain
spatial positions, a quasinormal mode expansion of the Green function is also
possible \cite{SH1}--\cite{SH2}.

An important expression relating the Green function and modal summation is
obtained by integrating (\ref{GTE}) with respect to frequency and using the
Sokhotski--Plemelj (SP) identity%
\begin{equation}
\lim_{\varepsilon\rightarrow0^{+}}\frac{1}{x\pm i\varepsilon}=\text{PV}\left(
\frac{1}{x}\right)  \mp i\pi\delta\left(  x\right)  ,
\end{equation}
leading to
\begin{equation}
\frac{1}{\pi}\int_{0}^{\infty}d\omega\frac{\omega^{2}}{c^{2}}\operatorname{Im}%
\mathbf{G}\left(  \mathbf{r},\mathbf{r}^{\prime},\omega\right)  =%
{\displaystyle\sum\nolimits_{\mathbf{k}}}
\frac{\omega_{\mathbf{k}}}{2}\mathbf{E}_{\mathbf{k}}\left(  \mathbf{r}%
,\omega_{\mathbf{k}}\right)  \mathbf{E}_{\mathbf{k}}^{\ast}\left(
\mathbf{r}^{\prime},\omega_{\mathbf{k}}\right)  .\label{p1}%
\end{equation}
This is the key relationship that allows converting between the Langevin noise approach and normal-mode QED,
and will be needed in the following. Since the case $\mathbf{r}=\mathbf{r}%
^{\prime}$ is often needed in field-atom interactions, it is worth noting that
in the event of material loss at point $\mathbf{r}$ ($\operatorname{Im}\left(
\mathbf{\varepsilon}\left(  \mathbf{r},\omega_{\lambda}\right)  \right)  >0$),
$\operatorname{Im}\mathbf{G}\left(  \mathbf{r},\mathbf{r},\omega\right)
\rightarrow\infty$, which is not seen with the transverse Green
function/transverse mode expansion.

\subsection{Langevin Noise Approach}

The Langevin noise approach is developed in detail in \cite{Welsch0}-\cite{B2}, and here we merely
use the main results as needed. We now allow a dispersive absorbing
(complex-valued) permittivity, with causality requiring $\mathbf{\varepsilon
}\left(  \mathbf{r,-}\omega\right)  =\mathbf{\varepsilon}^{\ast}\left(
\mathbf{r,}\omega^{\ast}\right)  $. For the Green function, $\mathbf{G}^{\ast
}\mathbf{(r},\mathbf{r}^{\prime},\omega\mathbf{)}=\mathbf{G(r},\mathbf{r}%
^{\prime},-\omega^{\ast}\mathbf{)}$, and we impose the condition
\begin{equation}
\mathbf{G(r},\mathbf{r}^{\prime},\omega\mathbf{)}\mathbf{\rightarrow0}\text{
for }\left\vert \mathbf{r-r}^{\prime}\right\vert \rightarrow\infty
\end{equation}
associated with some material absorption. This is an often-overlooked
requirement, which is discussed further in Section \ref{Comments}.

The electric field operator in the Schr\"{o}dinger picture is%
\begin{equation}
\widehat{\mathbf{E}}\left(  \mathbf{r}\right)  ^{\text{LNA}}=\int_{0}^{\infty
}d\omega_{\lambda}\ \widehat{\mathbf{E}}\left(  \mathbf{r},\omega_{\lambda
}\right)  +\text{H.c.},\label{OLNA}%
\end{equation}
where $\omega_{\lambda}$ is a continuum modal frequency (not a Fourier
transform frequency), with%
\begin{align}
\widehat{\mathbf{E}}\left(  \mathbf{r},\omega_{\lambda}\right)  =i\sqrt
{\frac{\hslash}{\pi\varepsilon_{0}}} &  \frac{\omega_{\lambda}^{2}}{c^{2}}\int
d^{3}\mathbf{r}^{\prime}\mathbf{G}\left(  \mathbf{r},\mathbf{r}^{\prime
},\omega_{\lambda}\right) \label{LNAO}\\
&  \cdot\sqrt{\operatorname{Im}\left(  \mathbf{\varepsilon}\left(
\mathbf{r}^{\prime},\omega_{\lambda}\right)  \right)  }\cdot\widehat
{\mathbf{f}}\left(  \mathbf{r}^{\prime},\omega_{\lambda}\right)  ,\nonumber
\end{align}
where $\widehat{\mathbf{f}},\widehat{\mathbf{f}}^{\dag}$ are canonically
conjugate field variables, which are continuum bosonic operator--valued
vectors of the combined matter-field system that satisfy
\begin{align}
\left[  \widehat{f}_{k}\left(  \mathbf{r},\omega\right)  ,\widehat
{f}_{k^{\prime}}^{\dag}\left(  \mathbf{r}^{\prime},\omega^{\prime}\right)
\right]   &  =\delta_{kk^{\prime}}\delta\left(  \omega-\omega^{\prime}\right)
\delta\left(  \mathbf{r}-\mathbf{r}^{\prime}\right)  ,\\
\left[  \widehat{f}_{k}\left(  \mathbf{r},\omega\right)  ,\widehat
{f}_{k^{\prime}}\left(  \mathbf{r}^{\prime},\omega^{\prime}\right)  \right]
&  =\left[  \widehat{f}_{k}^{\dag}\left(  \mathbf{r},\omega\right)
,\widehat{f}_{k^{\prime}}^{\dag}\left(  \mathbf{r}^{\prime},\omega^{\prime
}\right)  \right]  =0.
\end{align}
Comparing the two approaches, $\int_{0}^{\infty}d\omega_{\lambda}$
$\widehat{\mathbf{f}}\left(  \mathbf{r},\omega_{\lambda}\right)  $ is seen to
be the continuous analog of $\sum_{\mathbf{k},s}\mathbf{E}_{\mathbf{k}}\left(
\mathbf{r}\right)  \widehat{a}_{\mathbf{k}s}$.

More complicated environments, including nonlocal, and nonreciprocal media,
have also been considered \cite{Buh}, \cite{RSW}-\cite{Force2}. The
conclusions described below hold for generally lossy, inhomogeneous,
nonreciprocal media

The free field-matter Hamiltonian is
\begin{equation}
\widehat{H}^{\text{LNA}}=\int_{0}^{\infty}d\omega\int d\mathbf{r\ }%
\hslash\omega\widehat{\mathbf{f}}^{\dag}\left(  \mathbf{r},\omega\right)
\cdot\widehat{\mathbf{f}}\left(  \mathbf{r},\omega\right)  ,
\end{equation}
which is analogous to (\ref{CMH}). Energy eigenstates of the free Hamiltonian
are compositions of $\left\vert n_{i}\left(  \mathbf{r},\omega_{\lambda
}\right)  \right\rangle $ (analogous to $\left\vert \left\{  n\right\}
\right\rangle $ in the cavity-mode case), which \thinspace indicates that the
$\lambda^{\text{th}}$ field mode of the nonuniform continuum is populated with
$n$ quanta, and that it is vector-valued with field component in the
$i^{\text{th}}$ direction. As a trivial example, the one-quanta states are
obtained from the ground state as
\begin{equation}
\left\vert 1_{i}\left(  \mathbf{r},\omega_{\lambda}\right)  \right\rangle
=\hat{f}_{i}^{\dagger}(\mathbf{r},\omega_{\lambda})\left\vert \left\{
0\right\}  \right\rangle .
\end{equation}

An important relation in developing Langevin noise approach formulations is the ``magic formula" \cite{Welsch1}%
\begin{align}
\frac{\omega^{2}}{c^{2}}\int d^{3}\mathrm{r}^{\prime} &  \operatorname{Im}%
\left(  \varepsilon\left(  \mathbf{r}^{\prime},\omega\right)  \right)
\boldsymbol{\mathrm{G}}(\mathbf{r},\mathbf{r}^{\prime},\omega)\cdot
\boldsymbol{\mathrm{G}}^{\text{\dag}}(\mathbf{r}_{0},\mathbf{r}^{\prime
},\omega)\label{magic1}\\
&  =\operatorname{Im}\boldsymbol{\mathrm{G}}(\mathbf{r},\mathbf{r}_{0}%
,\omega),\nonumber
\end{align}
generalized for tensor permittivity as \cite{Buh}, \cite{Hanson}%
\begin{align}
\frac{\omega^{2}}{c^{2}}\int d^{3}\mathrm{r}^{\prime\prime}%
\boldsymbol{\mathrm{G}}(\mathbf{r},\mathbf{r}^{\prime\prime},\omega) &
\cdot\boldsymbol{\mathrm{T}}(\mathbf{r}^{\prime\prime},\omega)\cdot
\boldsymbol{\mathrm{T}}^{\dagger}(\mathbf{r}^{\prime\prime},\omega
)\cdot\boldsymbol{\mathrm{G}}^{\dag}(\mathbf{r}^{\prime},\mathbf{r}%
^{\prime\prime},\omega)\nonumber\\
&  =\left(  \boldsymbol{\mathrm{G}}(\mathbf{r},\mathbf{r}^{\prime}%
,\omega)-\boldsymbol{\mathrm{G}}^{\dagger}(\mathbf{r}^{\prime},\mathbf{r}%
,\omega)\right)  /2i,\label{magic2}%
\end{align}
where $\boldsymbol{\mathrm{T}}(\mathbf{r},\omega)=\sqrt{\mathrm{Im}%
\boldsymbol{\varepsilon}(\mathbf{r},\omega)}$ (and valid for nonreciprocal
media using $\boldsymbol{\mathrm{T}}(\mathbf{r},\omega)\cdot
\boldsymbol{\mathrm{T}}^{\dagger}(\mathbf{r},\omega)=\frac{1}{2i}\left[
\boldsymbol{\varepsilon}(\mathbf{r},\omega)-\boldsymbol{\varepsilon}^{\dagger
}(\mathbf{r},\omega)\right]  $). The above integrals generally don't need to
be evaluated explicitly, but are used in the derivation of system equations;
their use removes $\operatorname{Im}\left(  \varepsilon\left(  \mathbf{r}%
^{\prime},\omega\right)  \right)  $ from the resulting equations, allowing the
lossless limit to be subsequently taken.

Furthermore, the correlation relation can be shown to be \cite{Buh}
\[
\left\langle 0|\mathbf{E}(\mathbf{r},\omega)\mathbf{E}^{\dagger}%
(\mathbf{r},\omega^{\prime})|0\right\rangle =\frac{\hbar k_{0}^{2}%
}{2\varepsilon_{0}^{2}}N\operatorname{Im}(\mathbf{G}(\mathbf{r},\mathbf{r}%
,\omega))\delta\left(  \omega-\omega^{\prime}\right)  ,
\]
where $N(\omega,T)=2/\left(  \mathrm{exp}(\hbar\omega/k_{B}T)-1\right)  $ for
negative frequencies and $N(\omega,T)=1+2/\left(  \mathrm{exp}(\hbar
\omega/k_{B}T)-1\right)  $ for positive frequencies, where $k_{B}$ is
Boltzmann's constant.

Conversion to the time-domain is achieved by changing to the Heisenberg
picture, where operators $\widehat{A}$ transform as $\widehat{A}_{H}\left(
t\right)  =e^{i\widehat{H}_{\mathrm{Sch}}t/\hslash}\widehat{A}_{\mathrm{Sch}%
}e^{-i\widehat{H}_{\mathrm{Sch}}t/\hslash}$, leading to
\begin{align}
&  \widehat{\mathbf{E}}\left(  \mathbf{r},t\right)  =\int_{0}^{\infty}%
d\omega_{\lambda}i\sqrt{\frac{\hslash}{\pi\varepsilon_{0}}}\frac
{\omega_{\lambda}^{2}}{c^{2}}\\
&  \times\int\mathbf{G}\left(  \mathbf{r},\mathbf{r}^{\prime},\omega_{\lambda
}\right)  \cdot\sqrt{\operatorname{Im}\left(  \mathbf{\varepsilon}\left(
\mathbf{r}^{\prime},\omega_{\lambda}\right)  \right)  }\cdot\widehat
{\mathbf{f}}\left(  \mathbf{r}^{\prime},\omega_{\lambda},t\right)
d^{3}\mathbf{r}^{\prime}+\text{H.c.}\nonumber
\end{align}

In summary, to compare the two methods, the normal-mode QED is the standard method
ubiquitous in quantum optics. It is a natural and convenient method to study
cavity-QED (e.g., Jaynes--Cummings models), nonclassical light, and
many-quanta correlations. It puts the system background (e.g., cavity) on a
similar footing as the system (e.g., an atom), both being modes/harmonic
oscillators. The Langevin noise approach is a system-bath approach which focuses attention on the
system (e.g., the atom) and, while rigorously accounting for the system
environment, the latter being relegated to the status of a bath. Although normal-mode QED can be complimented by system-bath decay operators which
approximately account for the non-Hermitian (outgoing and incoming) nature of
the cavity modes in real systems, the commutation rules assumed are formally
only valid for $Q\rightarrow\infty$, a restriction not needed for the Langevin noise approach. In
the Langevin noise approach, there is often some confusion about the integration limits
and the limit $\operatorname{Im}\left(  \varepsilon\left(  \mathbf{r}%
,\omega\right)  \right)  \rightarrow0$, discussed further 
in the following.

\section{Example I: Excited Atom Introduced into a Structured Reservoir --
Non-Markovian Weisskopf-Wigner Analysis\label{IEA}}

As a first example, in this section we consider introducing an excited-state
atom at $\mathbf{r}=\mathbf{r}_{0}$, $t=0$ into a structured reservoir
\cite{Force2}, comparing the normal-mode QED and Langevin noise approaches in the context of 3D
quantization in the limit $\operatorname{Im}\left(  \mathbf{\varepsilon
}\left(  \mathbf{r},\omega\right)  \right)  \rightarrow0$.

The Hamiltonian operator is
\begin{equation}
H=H^{\text{NMQED/LNA}}+\hbar\omega_{0}\hat{\sigma}_{+}{\sigma}_{-}%
-\hat{\mathbf{p}}\cdot\widehat{\boldsymbol{\mathrm{E}}}(\mathbf{r}_{0}),
\end{equation}
where $\hat{\sigma}_{\pm}$ are the canonically conjugate two-level atomic
operators ($\widehat{\sigma}_{+}=\left\vert e\right\rangle \left\langle
g\right\vert ,\ \widehat{\sigma}_{-}=\left\vert g\right\rangle \left\langle
e\right\vert =\widehat{\sigma}_{+}^{\dagger}$, with $\left\vert e\right\rangle
$ and $\left\vert g\right\rangle $ being the excited and ground atomic states,
respectively), and $\hat{\mathbf{p}}=\left(  \hat{\sigma}_{+}+\hat{\sigma}%
_{-}\right)  \mathbf{\gamma}$ is the dipole operator, where $\mathbf{\gamma}$
is the dipole operator matrix-element, assumed real-valued. The first term in
each case is the free Hamiltonian for the field modes (field-matter modes for
the Langevin noise approach), the second term is the free Hamiltonian for the dipole, and the last
term is the interaction term.

The equation of motion is%
\begin{equation}
\frac{d}{dt}\left\vert \psi\right\rangle =-\frac{i}{\hslash}H\left\vert
\psi\right\rangle ,\label{EOM}%
\end{equation}
and in each case the atom-field product states are
\begin{align}
\left\vert \psi\left(  t\right)  \right\rangle ^{\text{NMQED}} &  =c_{e}\left(
t\right)  \left\vert e,0\right\rangle +\sum_{\lambda}c_{\lambda}\left(
t\right)  \left\vert g,1_{\lambda}\right\rangle \label{WFCNMA}\\
\left\vert \psi\left(  t\right)  \right\rangle ^{\text{LNA}} &  =c_{e}\left(
t\right)  \left\vert e,0\right\rangle \label{WFLNA}\\
&  +\int d^{3}\mathbf{r}\int_{0}^{\infty}d\omega_{\lambda}c_{gi}\left(
\mathbf{r},\omega_{\lambda},t\right)  \left\vert g,1_{i}\left(  \mathbf{r}%
,\omega_{\lambda}\right)  \right\rangle ,\nonumber
\end{align}
where $\left\vert e,0\right\rangle \equiv\left\vert e\right\rangle
\otimes\left\vert \left\{  0\right\}  \right\rangle $ and\ $\left\vert
g,1_{i}\left(  \mathbf{r},\omega_{\lambda}\right)  \right\rangle
\equiv\left\vert g\right\rangle \otimes\left\vert \left\{  1_{i}\left(
\mathbf{r},\omega_{\lambda}\right)  \right\}  \right\rangle $. The interaction
Hamiltonian $\hat{\mathbf{p}}\cdot\widehat{\boldsymbol{\mathrm{E}}}%
(\mathbf{r}_{0})\sim\left(  \hat{\sigma}_{+}+\hat{\sigma}_{-}\right)  \left(
\hat{\boldsymbol{\mathrm{f}}}+\hat{\boldsymbol{\mathrm{f}}}^{\dagger}\right)
$ acting on the initial state $\left\vert e,0\right\rangle $ leads to an
infinite-dimensional Hilbert space of the set of states $A=\left\{  \left\vert
e,0\right\rangle ,\left\vert g,1\right\rangle ,\left\vert e,2\right\rangle
,\left\vert g,3\right\rangle ,\left\vert e,4\right\rangle ,...\right\}  $,
where the $n>1$ photons could be in the same or different field modes. Here,
we truncate the space to consist of \{$\left\vert e,0\right\rangle ,\left\vert
g,1\right\rangle $\}, which is equivalent to a rotating wave approximation
even when using the full interaction Hamiltonian.

For the normal-mode QED, plugging $\left\vert \psi\left(  t\right)  \right\rangle
^{\text{NMQED}}$ into the equation of motion and defining%
\begin{equation}
g_{\mathbf{k}}=\mathbf{\gamma}\cdot i\sqrt{\frac{\hslash\omega_{\mathbf{k}}%
}{2\varepsilon_{0}}}\mathbf{E}_{\mathbf{k}}\left(  \mathbf{r}_{0}\right)
,\label{g}%
\end{equation}
multiplying by $\left\langle e,0\right\vert $ and $\left\langle g,1_{\lambda
^{\prime}}\right\vert $, and discarding higher-order terms like $\widehat
{a}_{\mathbf{k}s}^{\dag}\left(  0\right)  \left\vert g,1_{\lambda
}\right\rangle \sim\left\vert g,2_{\lambda}\right\rangle $, leads to
\cite{QOB2}
\begin{align}
\frac{d}{dt}c_{e} &  =-ic_{e}\omega_{0}+\frac{i}{\hslash}\sum_{\lambda
}g_{\lambda}c_{\lambda}\\
\frac{d}{dt}c_{\lambda} &  =\frac{i}{\hslash}c_{e}g_{\lambda}^{\ast}%
-i\omega_{\lambda}c_{\lambda}.
\end{align}
Defining slowly-varying amplitudes $c_{es}\left(  t\right)  =c_{e}\left(
t\right)  e^{i\omega_{0}t}$ and $c_{\lambda s}\left(  t\right)  =c_{\lambda
}\left(  t\right)  e^{i\omega_{\lambda}t}$, where $\omega_{0}$ is the energy
level transition frequency, we have%
\begin{equation}
c_{\lambda s}\left(  t\right)  =\frac{i}{\hslash}g_{\lambda}^{\ast}\int
_{0}^{t}dt^{\prime}c_{es}\left(  t^{\prime}\right)  e^{i\left(  \omega
_{\lambda}-\omega_{0}\right)  t^{\prime}}%
\end{equation}
and so the population is obtained by solving the Volterra integral equation of the second kind
\begin{equation}
\frac{dc_{es}\left(  t\right)  }{dt}=\int_{0}^{t}D\left(  t,t^{\prime}\right)
c_{es}\left(  t^{\prime}\right)  dt^{\prime},\label{VIE}%
\end{equation}
with the kernel%
\begin{equation}
D^{\text{NMQED}}\left(  t,t^{\prime}\right)  =-\frac{1}{\hslash^{2}}%
\sum_{\lambda}\left\vert g_{\lambda}\right\vert ^{2}e^{-i\left(
\omega_{\lambda}-\omega_{0}\right)  \left(  t-t^{\prime}\right)  }.
\end{equation}
The Volterra integral equation has been widely utilized in quantum optics, see, e.g., \cite{BF1}%
-\cite{WW1}, and can accommodate non-Markovian processes. The procedure for
numerically solving the Volterra integral equation is shown in \cite{Force2},
\cite{NR}. The initial-value condition $c_{eo}\left(  0\right)  =1$ is
assumed, representing an initially-excited atom.

Repeating the same procedure for the Langevin noise approach (details are in
\cite{Force2}) leads to (\ref{VIE}) where \cite{Trung}, \cite{QED}%
\begin{align}
D^{\text{LNA}}\left(  t,t^{\prime}\right)  = & \label{DNLA}\\
-\frac{1}{\hslash\pi\varepsilon_{0}}\int_{0}^{\infty} &  d\omega_{\lambda
}\ \frac{\omega_{\lambda}^{2}}{c^{2}}\mathbf{\gamma}\cdot\operatorname{Im}%
\mathbf{G}(\mathbf{r}_{0},\mathbf{r}_{0},\omega_{\lambda})\cdot\mathbf{\gamma
}\nonumber\\
&  \times e^{-i\left(  \omega_{\lambda}-\omega_{0}\right)  \left(
t-t^{\prime}\right)  }\doteqdot D^{\text{NMQED}}\left(  t,t^{\prime}\right)
,\nonumber
\end{align}
using (\ref{magic1}), (\ref{g}) and (\ref{p1}), where $\doteqdot$ indicates
equality in the lossless limit of the Langevin noise approach formulation (that is, when
(\ref{p1}) holds). The term $\sqrt{\operatorname{Im}\left(
\mathbf{\varepsilon}\left(  \mathbf{r}^{\prime},\omega_{\lambda}\right)
\right)  }$ does not appear in the expression for $D^{\text{LNA}}$. Since the
Langevin noise approach can accommodate generally lossy, dispersive media, the Langevin noise approach
approach exactly recovers the normal-mode QED as a special case. There is no need to
explicitly take the limit as $\operatorname{Im}\left\{  \mathbf{\varepsilon
}\left(  \mathbf{r},\omega\right)  \right\}  \rightarrow\mathbf{0}$, one
merely computes the Green function assuming lossless media. This is discussed
further in Section \ref{Comments}. The Langevin noise approach also applies to open systems, where
the Green function accounts for the infinite space. The vacuum limit is
obtained merely by using the vacuum Green function.

To recover the familiar Markov result, setting $c_{es}\left(  t^{\prime
}\right)  =c_{es}\left(  t\right)  $, and using the SP identity $\int
_{0}^{\infty}e^{\pm i\left(  \omega-\omega_{0}\right)  \tau}d\tau=\pi
\delta\left(  \omega-\omega_{0}\right)  \pm i$PV$\left(  \frac{1}%
{\omega-\omega_{0}}\right)  $, (\ref{VIE}) can be solved as
\begin{equation}
c_{es}\left(  t\right)  =c_{es}\left(  0\right)  e^{-\Gamma\frac{1}{2}%
t}e^{i\delta t},\label{e3}%
\end{equation}
and the probability of excited state occupation is $P\left(  t\right)
=\left\vert c_{es}\left(  t\right)  \right\vert ^{2}=\left\vert c_{es}\left(
0\right)  \right\vert ^{2}e^{-\Gamma t}$. In (\ref{e3}),
\begin{align}
\Gamma &  =2\frac{\pi}{\hslash\pi\varepsilon_{0}}\frac{\omega_{0}^{2}}{c^{2}%
}\mathbf{\gamma}\cdot\operatorname{Im}\mathbf{G}(\mathbf{r}_{0},\mathbf{r}%
_{0},\omega_{\lambda})\cdot\mathbf{\gamma},\\
\delta &  =\frac{1}{\hslash\pi\varepsilon_{0}}\text{PV}\int_{0}^{\infty
}d\omega_{\lambda}\ \frac{\omega_{\lambda}^{2}}{c^{2}}\frac{\mathbf{\gamma
}\cdot\operatorname{Im}\mathbf{G}(\mathbf{r}_{0},\mathbf{r}_{0},\omega
_{\lambda})\cdot\mathbf{\gamma}}{\left(  \omega_{\lambda}-\omega_{0}\right)
},
\end{align}
where $\Gamma$ is the usual decay rate \cite{Nov}, and for vacuum,
$\Gamma^{\text{vac}}=\gamma^{2}\omega_{0}^{3}/\pi\varepsilon_{0}\hslash c^{3}%
$. Note that here we start with the Green function and obtain the normal mode
result, whereas in \cite{Wubs2} they start with the normal modes and obtain
the Green function (albeit for the lossy case).

\section{Example II: Driven Atom in a Structured Reservoir -- Density Operator
Analysis\label{IEA2}}

As a second example, we consider an atom in a structured reservoir under the
action of an external pump. The derivation follows the familiar route
\cite{OQS}, and, for the Langevin noise approach details are available in \cite{Hanson}. The
resulting Schr\"{o}dinger picture master equation (ME) is, under the Born and
Markov approximations,
\begin{align}
&  \frac{d}{dt}\rho\left(  t\right)  =-\frac{i}{\hslash}\left[  \widehat
{H}_{\text{S}},\rho\left(  t\right)  \right]  -\int_{0}^{t}d\tau\left(
J_{ph}^{n+1}\left(  \tau\right)  \widehat{\sigma}_{+}\widehat{\sigma}%
_{-}\left(  -\tau\right)  \rho\left(  t\right)  \right. \nonumber\\
&  -J_{ph}^{n+1}\left(  \tau\right)  \widehat{\sigma}_{-}\left(  -\tau\right)
\rho\left(  t\right)  \widehat{\sigma}_{+}-J_{ph}^{n+1}\left(  -\tau\right)
\widehat{\sigma}_{-}\rho\left(  t\right)  \widehat{\sigma}_{+}\left(
-\tau\right) \nonumber\\
&  +J_{ph}^{n+1}\left(  -\tau\right)  \rho\left(  t\right)  \widehat{\sigma
}_{+}\left(  -\tau\right)  \widehat{\sigma}_{-}-J_{ph}^{n}\left(
-\tau\right)  \widehat{\sigma}_{-}\widehat{\sigma}_{+}\left(  -\tau\right)
\rho\left(  t\right) \nonumber\\
&  +J_{ph}^{n}\left(  -\tau\right)  \widehat{\sigma}_{+}\left(  -\tau\right)
\rho\left(  t\right)  \widehat{\sigma}_{-}+J_{ph}^{n}\left(  \tau\right)
\widehat{\sigma}_{+}\rho\left(  t\right)  \widehat{\sigma}_{-}\left(
-\tau\right) \nonumber\\
&  -J_{ph}^{n}\left(  \tau\right)  \rho\left(  t\right)  \widehat{\sigma}%
_{-}\left(  -\tau\right)  \widehat{\sigma}_{+}.\label{MFE}%
\end{align}
where $H_{\text{S}}=\hslash\left(  \omega_{d}-\omega_{L}\right)
\widehat{\sigma}^{+}\widehat{\sigma}^{-}+\frac{\hslash\Omega}{2}\left(
\sigma^{+}+\sigma^{-}\right)  $. For the normal-mode QED,%
\begin{align}
J_{ph}^{n+1}\left(  \tau\right)   &  =%
{\displaystyle\sum\nolimits_{\mathbf{k}}}
J_{\mathbf{k}}\left(  \overline{n}\left(  \omega_{\mathbf{k}}\right)
+1\right)  e^{-i\left(  \omega_{\mathbf{k}}-\omega_{L}\right)  \tau}\\
J_{ph}^{n}\left(  \tau\right)   &  =%
{\displaystyle\sum\nolimits_{\mathbf{k}}}
J_{\mathbf{k}}\overline{n}\left(  \omega_{\mathbf{k}}\right)  e^{-i\left(
\omega_{\mathbf{k}}-\omega_{L}\right)  \tau}\\
J_{\mathbf{k}} &  =\frac{\omega_{\mathbf{k}}}{2\hslash\varepsilon_{0}%
}\mathbf{\gamma}\cdot\mathbf{E}_{\mathbf{k}}\left(  \mathbf{r}\right)
\mathbf{E}_{\mathbf{k}}^{\ast}\left(  \mathbf{r}\right)  \cdot\mathbf{\gamma,}%
\end{align}
and for the Langevin noise approach,
\begin{align}
J_{ph}^{n+1}\left(  \tau\right)   &  =\int_{0}^{\infty}d\omega J_{ph}\left(
\omega\right)  \left(  \overline{n}\left(  \omega\right)  +1\right)
e^{-i\left(  \omega-\omega_{L}\right)  \tau}\\
J_{ph}^{n}\left(  \tau\right)   &  =\int_{0}^{\infty}d\omega J_{ph}\left(
\omega\right)  \overline{n}\left(  \omega\right)  e^{-i\left(  \omega
-\omega_{L}\right)  \tau}\\
J_{ph}\left(  \omega\right)   &  =\frac{\omega^{2}}{c^{2}}\frac{\mathbf{\gamma
}\cdot\operatorname{Im}\left(  \mathbf{G}\left(  \mathbf{r},\mathbf{r}%
,\omega\right)  \right)  \cdot\mathbf{\gamma}}{\pi\hslash\varepsilon_{0}}%
\end{align}
and, where $\overline{n}$ is the average number of thermal photons,
$\overline{n}=\left(  e^{\frac{\hslash\omega}{k_{B}T}}-1\right)  ^{-1}$.

Using (\ref{p1}), it is easy to show that%
\begin{equation}%
{\displaystyle\sum\nolimits_{\mathbf{k}}}
J_{\mathbf{k}}e^{-i\omega_{\mathbf{k}}\tau}=\int_{0}^{\infty}d\omega J\left(
\omega\right)  e^{-i\omega\tau},
\end{equation}
and, thus,
\begin{equation}
\frac{d\rho\left(  t\right)  ^{\text{LNA}}}{dt}\doteqdot\frac{d\rho\left(
t\right)  ^{\text{NMQED}}}{dt},
\end{equation}
and the system evolution is the same for both approaches.

As a special case, if we set $\overline{n}=0$ and turn off the pump,
$H_{\text{S}}=\hslash\omega_{d}\widehat{\sigma}^{+}\widehat{\sigma}^{-}$, in
which case $\widehat{\sigma}_{\mp}\left(  -\tau\right)  =\widehat{\sigma}%
_{\mp}e^{\pm i\omega_{d}\tau}$, we obtain the familiar ME for a single atom
interacting with its environment,
\begin{align}
\frac{d}{dt}\rho &  =-i\left(  \omega_{d}-\Delta_{d}\right)  \left[
\widehat{\sigma}^{+}\widehat{\sigma}^{-},\rho\left(  t\right)  \right] \\
&  +\frac{\gamma\left(  \omega_{d}\right)  }{2}\left(  2\widehat{\sigma}%
_{-}\rho\left(  t\right)  \widehat{\sigma}_{+}-\widehat{\sigma}_{+}%
\widehat{\sigma}_{-}\rho\left(  t\right)  -\rho\left(  t\right)
\widehat{\sigma}_{+}\widehat{\sigma}_{-}\right)  ,
\end{align}
where we used the SP identity and where $\gamma\left(  \omega_{d}\right)
=2\pi J\left(  \omega_{d}\right)  $, $\Delta_{d}=\frac{1}{\hslash^{2}}$%
PV$\int_{0}^{\infty}d\omega J\left(  \omega\right)  /\left(  \omega-\omega
_{d}\right)  $. The ME for a multi-atom system, allowing for, e.g., the study
of entanglement, is also the same for the normal-mode QED and Langevin noise approaches.

\section{Comments on the Connection Between Normal-Mode QED and Langevin Noise Approaches, and
Validity of the Langevin Noise Approach\label{Comments}}

Normal-mode QED is well-founded mathematically, based on canonical quantization and
completeness of the eigenfunctions of self-adjoint operators \footnote{From a
strict mathematical perspective, self-adjointness is actually not enough, but
SA together with having a compact inverse does guarantee the validity of
eigenfunction expansions \cite{OT}. This condition is satisfied by typical
differential equations arising in electromagnetics and quantum mechanical
problems, which have compact integral inverse operators.}. Much of quantum
optics is based on electric field operators of the form (\ref{EFO1}%
)-(\ref{EFO2}) using planewave eigenfunctions (\ref{PWEF}) (including
microscopic models). As more complicated environments have been considered,
the eigenfunctions based on (\ref{efe}) have been used. However, all of the
aforementioned eigenfunctions only form complete sets in limited settings
(closed cavities, usually lossless, dispersionless materials), where material
parameters are represented by Hermitian (self-adjoint) tensors. Note that
completeness is important, not only for (\ref{p1}), but also for validity of
the operators (\ref{EFO1})-(\ref{EFO2}), which are also eigenfunction expansions.

Two comments are important: 1) Some level of loss must be maintained in the
system when using the operator (\ref{OLNA})-(\ref{LNAO}); it is impermissible
to let $\operatorname{Im}\left(  \mathbf{\varepsilon}\left(  \mathbf{r}%
,\omega_{\lambda}\right)  \right)  \rightarrow0$ until after that term drops
out from the formulation, typically after using (\ref{magic1}) or
(\ref{magic2}). One can not take this limit in the operator (\ref{OLNA}%
)-(\ref{LNAO}). 2) If in Fig. 1 $\varepsilon_{\text{bulk}}$ is lossless, then
it is also impermissible to let the size of the region of interest shrink to
zero to implement the vacuum limit (i.e., $\Omega_{1}\rightarrow0$ in Fig. 1),
until after using (\ref{magic1}) or (\ref{magic2}), after which the Green
function is merely the vacuum Green function for the cavity or open space (if
$\varepsilon_{\text{bulk}}$ is lossy, than one can allow the limit $\Omega
_{1}\rightarrow0$ at the onset). In the presented examples, using (\ref{p1}),
the Langevin noise approach reduces to the normal-mode QED result for closed cavities; alternatively, using
(\ref{p1}), the normal-mode QED result can be generalized to involve the Green function,
allowing cavities with lossy, dispersive materials to be considered, and even
open geometries. However, this is not a general result (i.e., this does not
universally hold).

In a practical sense, lossless materials don't exist, aside from vacuum.
Therefore, it is not unreasonable to consider space to be filled with a
background medium having perhaps $\operatorname{Re}\left(  \varepsilon\right)
\simeq1$ and $\operatorname{Im}\left(  \varepsilon\right)  >0$, into which the
actual structure of interest is placed, as depicted in Fig. 1. The Green
function accounts for the entire permittivity $\mathbf{\varepsilon}\left(
\mathbf{r},\omega\right)  $, including the background, and after
$\operatorname{Im}\left(  \mathbf{\varepsilon}\left(  \mathbf{r}%
,\omega\right)  \right)  $ is removed from the formulation using
(\ref{magic1})-(\ref{magic2}) and only the Green function remains, one can
consider lossless materials.

\subsection{Lossless Limit of the ``Magic Formula" (\ref{magic1})}

The connection between the normal-mode QED and the Langevin noise approach is established by virtue of the conversion formula~(\ref{p1})---showing that normal-mode QED is a special case of the Langevin noise approach in the lossless limit. However, the explicit presence of the factor $\sqrt{\operatorname{Im}\left(  \mathbf{\varepsilon}\left(
\mathbf{r},\omega\right)  \right)  }$ in the field expansion~(\ref{LNAO}) indicates that this limit has to be understood in a strict sense as a mathematical limiting procedure where $\sqrt{\operatorname{Im}\left(  \mathbf{\varepsilon}\left(
\mathbf{r},\omega\right)  \right)  }\to 0$ while $\sqrt{\operatorname{Im}\left(  \mathbf{\varepsilon}\left(
\mathbf{r},\omega\right)  \right)  }> 0$. In fact, the presence of $\sqrt{\operatorname{Im}\left(  \mathbf{\varepsilon}\left(\mathbf{r},\omega\right)  \right)  }$ in the field explansion is an artifact of normalising the bosonic canonically conjugate field variables and is avoided if one instead works with the noise polarisation. 

In either case, after evaluating operator dynamics or taking quantum expectation values, one typically arrives at the left hand side of the integral relation~(\ref{magic1}). The right hand side of this formula is obviously finite in the above-defined lossless limit $\operatorname{Im}\epsilon(\mathbf{r},\omega)\to 0+$. At first glance, the left hand side seems to vanish in this limit due to the presence of the factor $\operatorname{Im}\left(  \mathbf{\varepsilon}\left(\mathbf{r}',\omega\right) \right)$. However, this conclusion is premature as a careful evaluation of the spatial integral will reveal a factor cancelling $\operatorname{Im}\left(  \mathbf{\varepsilon}\left(\mathbf{r}',\omega\right) \right)$, so that the limit may be taken to give the same result as the right hand side of the equation.

To illustrate this, consider the case of a bulk medium with permittivity $\varepsilon\left(\omega\right)=\varepsilon_\mathrm{R}\left(\omega\right)+\mi\delta$ with $\varepsilon_\mathrm{R}$ real. The respective Green tensor is given by
\begin{multline}
\label{B10}
\tens{G}^{(0)}(\vec{r},\vec{r}',\omega)
=-\frac{1}{3k^2}\,\bm{\delta}(\vec{\rho})
 -\frac{\me^{\mi k\rho}}{4\pi k^2\rho^3}
 \bigl\{\bigl[1-\mi k\rho-(k\rho)^2\bigl]\tens{I}\\
-\bigl[3-3\mi k\rho-(k\rho)^2\bigr]
\vec{e}_\rho\tprod\vec{e}_\rho\bigr\}
\end{multline}
with \mbox{$\vec{\rho}=\vec{r}-\vec{r}'$}; \mbox{$\rho=|\vec{\rho}|$};
\mbox{$\vec{e}_\rho=\vec{\rho}/\rho$}, and $k=\sqrt{\varepsilon\left(\omega\right)}\omega/c$ such that $\operatorname{Im}k>0$. In the limit $\vec{r}_0\to\vec{r}$, $\vec{r}_0\neq\vec{r}$, we hence have
\begin{equation}
\tens{G}^{(0)}(\vec{r},\vec{r}',\omega)\sprod\tens{G}^{(0)\dagger}(\vec{r}_0,\vec{r}',\omega)
\propto\me^{-2\delta\rho}.
\end{equation}
To leading order in $\delta$, this implies
\begin{equation}
\int\dif^3 \mathrm{r'}\,\tens{G}^{(0)}(\vec{r},\vec{r}',\omega)\sprod\tens{G}^{(0)\dagger}(\vec{r}_0,\vec{r}',\omega)=\operatorname{O}(1/\delta),
\end{equation} 
so that 
\begin{equation}
\int\dif^3 \mathrm{r'}\,\operatorname{Im}\varepsilon\left(\omega\right)\tens{G}^{(0)}(\vec{r},\vec{r}',\omega)\sprod\tens{G}^{(0)\dagger}(\vec{r}_0,\vec{r}',\omega)
\end{equation} 
remains finite in the limit $\operatorname{Im}\varepsilon\left(\omega\right)\equiv\delta\to0+$.
In App.~\ref{AppA}, we explicitly demonstrate the validity of the integral relation~(\ref{magic1}) in the lossless limit for the more general case of arbitrary~$\vec{r},\vec{r}_0$.

An alternative way to establish contact with the nonabosorbing case was suggested in Ref.~\cite{AD}. Here, the region of interest is surrounded by a strictly lossless region $\operatorname{Im}\varepsilon\left(\omega\right)=0$ at infinity (or sufficiently far, respectivly). It was shown that under such conditions the integral relation (\ref{magic1}) has an additional term
\begin{align}
\frac{\omega^{2}}{c^{2}}\int_{\Omega}d^{3}\mathbf{r}^{\prime} &
\operatorname{Im}\left(  \varepsilon\left(  \mathbf{r}^{\prime},\omega\right)
\right)  \boldsymbol{\mathrm{G}}(\mathbf{r},\mathbf{r}^{\prime},\omega
)\cdot\boldsymbol{\mathrm{G}}^{\text{\dag}}(\mathbf{r}_{0},\mathbf{r}^{\prime
},\omega)\label{magic3}\\
&  +%
{\displaystyle\oint\nolimits_{\Sigma}}
d^{2}\mathbf{r}^{\prime}\mathbf{F}\left(  \mathbf{r}^{\prime},\mathbf{r}%
,\mathbf{r}_{0}\right)  =\operatorname{Im}\boldsymbol{\mathrm{G}}%
(\mathbf{r},\mathbf{r}_{0},\omega),\nonumber
\end{align}
where%
\begin{align}%
{\displaystyle\oint\nolimits_{\Sigma}}
d^{2}\mathbf{r}^{\prime}\mathbf{F}\left(  \mathbf{r}^{\prime},\mathbf{r}%
,\mathbf{r}_{0}\right)   &  =\frac{\omega}{c}\sqrt{\varepsilon_{\text{bulk}}}%
{\displaystyle\oint\nolimits_{\Sigma}}
d^{2}\mathbf{r}^{\prime}\mathbf{G}^{\text{T}}\left(  \mathbf{r}^{\prime
},\mathbf{r}\right) \label{magic4}\\
&  \cdot\mathbf{R}\times\mathbf{R}\times\mathbf{G}^{\ast}\left(
\mathbf{r}^{\prime},\mathbf{r}_{0}\right)  \text{,}\nonumber
\end{align}
and $\Sigma$ is the bounding surface that is far from the system in
question. In the event of an absorbing (perhaps limitingly-low-loss)
background medium $\varepsilon_{\text{bulk}}$, the Green tensor vanishes on
$\Sigma$
and the surface contribution vanishes accordingly.
This is commensurate with the
requirement $\mathbf{G(r},\mathbf{r}^{\prime},\omega\mathbf{)}%
\mathbf{\rightarrow0}$ for $\left\vert \mathbf{r-r}^{\prime}\right\vert
\rightarrow\infty$. Thus, one must retain material absorption of the
background environment if (\ref{magic1}) or (\ref{magic2}) is to be used,
to ensure that no boundry contribution arises.
Physically, one could
argue that the assumption of a background environment without at least some
small amount of absorption is generally a fiction anyway, aside from perhaps
evacuated superconducting chambers. \ Alternatively, in Ref.~\cite{AD} it is shown
that implementing the developed scheme of replacing the missing free incident
field with polarization currents at infinity with a lossless interior region,
to bring the Langevin noise approach into accordance with the Huttner--Barnett result, and
including the boundary term one 
recovers
the usual Langevin noise approach.

\section{Conclusions}

The Langevin noise approach for quantization of macroscopic electromagnetics
for three-dimensional, inhomogeneous environments has been compared with the
usual normal mode quantization in quantum optics. The conditions of validity
of the normal mode expansion were discussed, and it was shown using several
examples that the Langevin noise approach reduces exactly to the normal mode expansion formulation
in the lossless limit. Conditions on applying the Langevin noise approach to finite structures
were also discussed. 

\section*{Acknowledgments}

The author gratefully acknowledge discussions with Stephen Hughes. 
This work was supported by the German Research Foundation
(DFG, Grant BU 1803/3-1).

\appendix

\section{Lossless limit for the bulk case}
\label{AppA}

In this appendix we explicitly show that the ``magic formula" in Eq.~\eqref{magic1} holds also in the limit of lossless media for the case of a single bulk dielectric material described by $\epsilon(\vec{r},\omega) =  \epsilon(\omega)$. This means we show that 
\begin{multline} \label{magicApp}
\lim_{\mathrm{Im}[\epsilon(\omega)]\to 0+}\hspace{-0.1cm}\frac{\omega^{2}}{c^{2}}\hspace{-0.2cm}\int \hspace{-0.1cm}d^{3}\mathrm{r}^{\prime}   \operatorname{Im}%
\left(  \varepsilon\left(  \mathbf{r}^{\prime},\omega\right)  \right) \\ \times
\boldsymbol{\mathrm{G}}^{(0)}(\mathbf{r},\mathbf{r}^{\prime},\omega)\cdot
\boldsymbol{\mathrm{G}}^{(0)\ast}(\mathbf{r}^{\prime
},\mathbf{r}_{0},\omega)\\
  =\,\,\,\,\,\,\, \mathclap{\lim_{\mathrm{Im}[\epsilon(\omega)]\to 0+}} \,\,\,\,\,\,\operatorname{Im}\boldsymbol{\mathrm{G}}^{(0)}(\mathbf{r},\mathbf{r}_{0}%
,\omega).
\end{multline}
Here, $\boldsymbol{\mathrm{G}}^{(0)}$ is the bulk Green tensor and note, that compared to Eq.~\eqref{magic1} we have already used that the Green's tensor for bulk isotropic dielectric material obeys Onsager reciprocity, i.e. $G_{ij}^{(0)}(\vec{r},\vec{r}^\prime) = G_{ji}^{(0)}(\vec{r}^\prime,\vec{r})$. We will show that Eq.~\eqref{magicApp} holds by using the bulk Green tensor $\tens{G}^{(0)}$ in its $(2+1)$-dimensional decomposition \cite{B1}
\begin{multline} \label{eq:G0}
\tens{G}^{(0)}(\vec{r},\vec{r}_0, \omega) =
- \frac{1}{ k^2} \delta^3(\vec{r} - \vec{r}_0) \vec{e}_z \vec{e}_z  \\
 + \frac{i}{8\pi^2} \int \dif^2 k_\parallel \frac{\me^{i\vec{k}_\parallel \cdot (\vec{r}-\vec{r}_0)}}{k^\perp} 
 \sum_{\sigma=s,p} \Big[ \vec{e}_{\sigma+}\vec{e}_{\sigma+} \me^{i k^\perp(z-z_0)} \theta(z - z_0)  \\  
+\vec{e}_{\sigma-}\vec{e}_{\sigma-} \me^{-i k^\perp(z-z_0)} \theta(z_0 - z)\Big].
\end{multline}
Here, $k^\perp = \sqrt{k^2-k_\parallel^2}$ and $k= \sqrt{\epsilon(\omega)}\omega/c$ and we have defined the polarisation vectors 
\begin{align} \label{eq:Pol1}
\vec{e}_{p \pm}  = \frac{1}{k_\parallel}\left( \begin{array}{c}
k_y \\ -k_x \\ 0 
\end{array} \right), \quad
\vec{e}_{p \pm}  = \frac{1}{k} \left( \begin{array}{c}
k^\perp k_x/k_\parallel \\ k^\perp k_y/k_\parallel \\ k_\parallel
\end{array}\right).
\end{align}
Inserting the first term of the Green's tensor in Eq.~\eqref{eq:G0} into the left hand side of Eq.~\eqref{magicApp} one obtains
\begin{multline}
 \lim_{\mathrm{Im}[\epsilon(\omega)]\to 0+}\frac{1}{|k|^4} \mathrm{Im}[\epsilon(\omega)] \frac{\omega^2}{c^2}   \int \dif^3 r^\prime \, \delta^3(\vec{r}-\vec{r}^\prime)  \delta^3(\vec{r}^\prime-\vec{r}_0)  \\
=  \lim_{\mathrm{Im}[\epsilon(\omega)]\to 0+} \frac{1}{|k|^4} \mathrm{Im}[\epsilon(\omega)] \frac{\omega^2}{c^2}  \delta^3(\vec{r}-\vec{r}_0)  =  0 .
\end{multline}
For the terms of the left hand side of Eq.~\eqref{magicApp} consisting of the product of a first and a second term of the bulk Green's tensor in Eq.~\eqref{eq:G0} one finds 
\begin{multline} \label{eq.4}
\lim_{\mathrm{Im}[\epsilon(\omega)]\to 0+} \mathrm{Im}[\epsilon(\omega)] \frac{\omega^2}{c^2}  \frac{\mi}{8\pi^2|k|^2} \int \dif^2 k_\parallel \me^{\mi \vec{k}_\parallel(\vec{r}-\vec{r}_0)} k_\parallel \\
 \times \bigg\{ \theta(z-z_0)  \left[ \frac{\vec{e}_z \vec{e}_{p-}^\ast \me^{-\mi k^{\perp \ast}(z-z_0)} }{k k^{\perp \ast}}  - \vec{e}_{p+}  \vec{e}_z  \frac{\me^{\mi k^{\perp }(z-z_0)} }{k^\ast k^{\perp}} \right] \\
+ \theta(z_0-z) \left[    \frac{ \vec{e}_z \vec{e}_{p+}^\ast \me^{\mi k^{\perp \ast}(z-z_0)}}{k k^{\perp \ast}}  -  \vec{e}_{p-} \vec{e}_z \frac{\me^{-\mi k^{\perp}(z-z_0)} }{k^\ast k^{\perp}}  \right]\bigg\}.
\end{multline}
This term again vanishes in the limit of $\mathrm{Im}[\epsilon(\omega)] \to 0$. \\
Hence, we are left with the terms stemming from the second term of the Green's tensor in Eq.~\eqref{eq:G0} only. Inserting the second and third rows of Eq.~\eqref{eq:G0} into the left hand side of Eq.~\eqref{magicApp} one finds
\begin{widetext}
\begin{multline} \label{eq.1}
\lim_{\mathrm{Im}[\epsilon(\omega)]\to 0+}  \mathrm{Im}[\epsilon(\omega)] \frac{\omega^2}{c^2}  \frac{1}{16\pi^2}   \int_{-\infty}^\infty \hspace{-0.4cm}\dif z^\prime \int \!\! \dif^2k_\parallel \frac{\me^{\mi \vec{k}_\parallel \cdot (\vec{r}-\vec{r}_0)} }{|k^\perp|^2 } \sum_{\sigma } \left[ \vec{e}_{\sigma+}\vec{e}_{\sigma+} \me^{i k^\perp(z-z^\prime)} \theta(z - z^\prime)
+\vec{e}_{\sigma-}\vec{e}_{\sigma-} \me^{-i k^\perp(z-z^\prime)} \theta(z^\prime - z)\right] \\
\cdot \left[ \vec{e}_{\sigma-}^\ast\vec{e}_{\sigma-}^\ast \me^{-i k^{\perp \ast}(z^\prime-z_0)} \theta( z^\prime- z_0)
+\vec{e}_{\sigma+}^{\ast }\vec{e}_{\sigma+}^{\ast } \me^{i k^{\perp\ast}(z^\prime-z_0)} \theta(z_0 -z^\prime)\right].
\end{multline}
Here, we carried out the $\vec{r}^\prime_{\parallel} $ integral leading to a factor $\delta^2(\vec{k}_\parallel + \vec{k}_\parallel^\prime)$ which in turn has been used to perform the $k^\prime_\parallel$ integral. Finally, we also used 
\begin{align} \label{rel1}
\vec{e}_{\sigma\pm}(-k_\parallel)\vec{e}_{\sigma\pm}(-k_\parallel) & = \vec{e}_{\sigma\mp}(k_\parallel)\vec{e}_{\sigma\mp}(k_\parallel) , \\
\vec{e}_{\sigma\pm}\cdot\vec{e}_{\sigma^\prime\pm}^{\ast } & = \delta_{\sigma \sigma^\prime} ,\\
\vec{e}_{\sigma\pm}\cdot\vec{e}_{\sigma^\prime\mp}^{\ast } & \propto \delta_{\sigma \sigma^\prime}.
\end{align}
The remaining $z'$ integral can be carried out straight forwardly and some lengthy algebra shows that Eq.~\eqref{eq.1} can be further reduced to 
\begin{multline} \label{eq.2}
 \lim_{\mathrm{Im}[\epsilon(\omega)]\to 0+} \mathrm{Im}[\epsilon(\omega)] \frac{\omega^2}{c^2}  \frac{1}{16\pi^2}  \int \dif^2k_\parallel \frac{\me^{\mi \vec{k}_\parallel \cdot (\vec{r}-\vec{r}_0)} }{|k^\perp|^2 } \\
  \times \sum_{\sigma } \left\{ \frac{1}{2\mathrm{Im}[k^\perp]}\left[ \vec{e}_{\sigma+}\vec{e}_{\sigma+}^\ast \me^{i k^\perp(z-z_0)} \theta(z - z_0)
+\vec{e}_{\sigma-}\vec{e}_{\sigma-}^\ast \me^{-i k^\perp(z-z_0)} \theta(z_0 - z)  \right. \right. \\
\left. \left.  +     \vec{e}_{\sigma-}\vec{e}_{\sigma-}^\ast \me^{-i k^{\perp\ast}(z-z_0)} \theta(z - z_0)
+\vec{e}_{\sigma+}\vec{e}_{\sigma+}^\ast \me^{+i k^{\perp\ast}(z-z_0)} \theta(z_0 - z)     \right] \right. \\
- \left.  \frac{\mi \vec{e}_{\sigma-}\cdot \vec{e}_{\sigma+}^\ast}{2 \mathrm{Re}[k^\perp]} \left[ \vec{e}_{\sigma+}\vec{e}_{\sigma-}^\ast \me^{i k^{\perp }(z-z_0)} \theta(z - z_0)
+\vec{e}_{\sigma-} \vec{e}_{\sigma+}^{\ast } \me^{i k^{\perp}(z_0-z)} \theta(z_0 - z)         \right. \right. \\
-  \left. \left. \vec{e}_{\sigma+}\vec{e}_{\sigma-}^\ast \me^{-i k^{\perp \ast}(z-z_0)} \theta(z - z_0)
-\vec{e}_{\sigma-} \vec{e}_{\sigma+}^{\ast } \me^{-i k^{\perp\ast}(z_0-z)} \theta(z_0 - z)   \right]    \right\}.
\end{multline}
To derive Eq.~\eqref{eq.2} we also used $\vec{e}_{\sigma-}\cdot \vec{e}_{\sigma+}^\ast = \vec{e}_{\sigma+}\cdot \vec{e}_{\sigma-}^\ast  $. Next, we rewrite
\begin{align}
\frac{1}{\mathrm{Im}[k^\perp]}  & =  \frac{|k^\perp|^2}{\mathrm{Im}[k^\perp]\mathrm{Re}[k^\perp]}\mathrm{Re}\bigg[\frac{1}{k^\perp}\bigg]   =  \frac{2|k^\perp|^2 c^2}{\mathrm{Im}[\epsilon(\omega)]\omega^2}\mathrm{Re}\bigg[\frac{1}{k^\perp}\bigg] , \\
\frac{1}{\mathrm{Re}[k^\perp]} &  =- \frac{|k^\perp|^2}{\mathrm{Re}[k^\perp]\mathrm{Im}[k^\perp]}\mathrm{Im}\bigg[\frac{1}{k^\perp}\bigg]  = - \frac{2|k^\perp|^2 c^2}{\mathrm{Im}[\epsilon(\omega)]\omega^2}\mathrm{Im}\bigg[\frac{1}{k^\perp}\bigg],
\end{align} 
in order to find that Eq.~\eqref{eq.2} is equivalent to 
\begin{multline} \label{eq.3}
  \lim_{ \epsilon(\omega) = 1+ \mi \delta \to 1}  \frac{1}{8\pi^2}  \int \dif^2k_\parallel \me^{\mi \vec{k}_\parallel \cdot (\vec{r}-\vec{r}_0)}\\
  \times \sum_{\sigma } \left\{  \mathrm{Re} \left[ \frac{1}{k^\perp}\right] \frac{1}{2} \left[ \vec{e}_{\sigma+}\vec{e}_{\sigma+}^\ast \me^{i k^\perp(z-z_0)} \theta(z - z_0)
+\vec{e}_{\sigma-}\vec{e}_{\sigma-}^\ast \me^{-i k^\perp(z-z_0)} \theta(z_0 - z)  \right. \right. \\
\left. \left.  +     \vec{e}_{\sigma-}\vec{e}_{\sigma-}^\ast \me^{-i k^{\perp\ast}(z-z_0)} \theta(z - z_0)
+\vec{e}_{\sigma+}\vec{e}_{\sigma+}^\ast \me^{+i k^{\perp\ast}(z-z_0)} \theta(z_0 - z)     \right] \right. \\
- \left.   \vec{e}_{\sigma-}\cdot \vec{e}_{\sigma+}^\ast \mathrm{Im}\left[ \frac{1}{k^\perp}\right] \frac{1}{2\mi} \left[ \vec{e}_{\sigma+}\vec{e}_{\sigma-}^\ast \me^{i k^{\perp }(z-z_0)} \theta(z - z_0)
+\vec{e}_{\sigma-} \vec{e}_{\sigma+}^{\ast } \me^{i k^{\perp}(z_0-z)} \theta(z_0 - z)         \right. \right. \\
-  \left. \left. \vec{e}_{\sigma+}\vec{e}_{\sigma-}^\ast \me^{-i k^{\perp \ast}(z-z_0)} \theta(z - z_0)
-\vec{e}_{\sigma-} \vec{e}_{\sigma+}^{\ast } \me^{-i k^{\perp\ast}(z_0-z)} \theta(z_0 - z)   \right]    \right\}.
\end{multline}
This was the crucial step, since the factor $\mathrm{Im}[\epsilon(\omega)]$ was cancelled meaning that now we are ready to take the limit $\mathrm{Im}[\epsilon(\omega)]\to 0 $. In this limit we find that $k\in \mathbb{R}$ which also leads to the fact that $k^\perp = \sqrt{k^2-k_\parallel^2 }$ is either real or purely imaginary depending on whether $k_\parallel <k $ or $k_\parallel > k$, respectively. This way we find that in the second and third row of Eq.~\eqref{eq.3} we can use
\begin{align}
\vec{e}_{\sigma\pm }^\ast = \vec{e}_{\sigma\pm } \quad \mathrm{if}\,\,\, k^\perp, k \in \mathbb{R};
\end{align}
whereas in the third and fourth row of Eq.~\eqref{eq.3} we have
\begin{align}
\vec{e}_{\sigma\pm }^\ast = \vec{e}_{\sigma\mp } \quad \mathrm{if}\,\,\, k^\perp \in \mi\mathbb{R}, k \in \mathbb{R}.
\end{align}
Since $\vec{e}_{\sigma\pm } \cdot \vec{e}_{\sigma^\prime \pm } = \delta_{\sigma \sigma^\prime}$, and using Eq.~\eqref{rel1} again we finally find that Eq.~\eqref{eq.3} can be further reduced to 
\begin{multline} \label{eq.Toshow}
 \frac{1}{8\pi^2} \int \dif^2 k_\parallel \me^{i\vec{k}_\parallel \cdot (\vec{r}-\vec{r}_0)}
 \sum_{\sigma=s,p}\\
 \times \left\{ \mathrm{Re}\left[\frac{1}{k^\perp} \right]\frac{1}{2} \Big[ \vec{e}_{\sigma+}\vec{e}_{\sigma+} \me^{i k^\perp(z-z_0)} \theta(z - z_0)
+\vec{e}_{\sigma-}\vec{e}_{\sigma-} \me^{-i k^\perp(z-z_0)} \theta(z_0 - z) + \mathrm{c.c.}(k_\parallel \to -k_\parallel )\Big] \right.  \\
\left.  - \mathrm{Im}\left[\frac{1}{k^\perp} \right]\frac{1}{2i} \Big[ \vec{e}_{\sigma+}\vec{e}_{\sigma+} \me^{i k^\perp(z-z_0)} \theta(z - z_0)
+\vec{e}_{\sigma-}\vec{e}_{\sigma-} \me^{-i k^\perp(z-z_0)} \theta(z_0 - z) - \mathrm{c.c.}(k_\parallel \to -k_\parallel )\Big]    \right\}.
\end{multline}
Here, $ \mathrm{c.c.}(k_\parallel \to -k_\parallel )$ denotes adding the complex conjugate of the preceding term which has also been subject to the replacement $k_\parallel \to -k_\parallel $. Equation~\eqref{eq.Toshow} is equivalent to $  \lim_{\mathrm{Im}[\epsilon(\omega)]\to 0+} \mathrm{Im} \tens{G}^{(0)}(\vec{r},\vec{r}^\prime, \omega )$ [cf.~Eq.~\eqref{eq:G0}] as desired.

\end{widetext}

\end{document}